
\magnification=1200
\voffset=0 true mm
\hoffset=0 true in
\hsize=6.5 true in
\vsize=8.5 true in
\normalbaselineskip=13pt
\def\doublespace{\baselineskip=20pt plus 3pt\message{double space}}
\def\singlespace{\baselineskip=13pt\message{single space}}
\let\spacing=\singlespace
\parindent=1.0 true cm



\newcount\equationumber \newcount\sectionumber 
\sectionumber=1 \equationumber=1               
\def\setsection{\global\advance\sectionumber by1 \equationumber=1} 
\def\numbe{{{\number\sectionumber}{.}\number\equationumber}
                            \global\advance\equationumber by1}
\def\numberit{\eqno{(\number\equationumber)} \global\advance\equationumber by1}
%
\def\numberal{(\number\equationumber)\global\advance\equationumber by1}
%
%
%
%


\def\ccf#1{\,\vcenter{\normalbaselines
    \ialign{\hfil$##$\hfil&&$\>\hfil ##$\hfil\crcr
      \mathstrut\crcr\noalign{\kern-\baselineskip}
      #1\crcr\mathstrut\crcr\noalign{\kern-\baselineskip}}}\,}
\def\scf#1{\,\vcenter{\baselineskip=9pt
    \ialign{\hfil$##$\hfil&&$\>\hfil ##$\hfil\crcr
      \vphantom(\crcr\noalign{\kern-\baselineskip}
      #1\crcr\mathstrut\crcr\noalign{\kern-\baselineskip}}}\,}

\def\small3j#1#2#3#4#5#6{\def\st{\scriptstyle} 
   \bigl(\scf{\st#1&\st#2&\st#3\cr
           \st#4&\st#5&\st#6\cr} \bigr)}


\def\ref#1{$^{#1)}$}    


\def\upa#1{\raise 1pt\hbox{\sevenrm #1}}
\def\dna#1{\lower 1pt\hbox{\sevenrm #1}}
\def\dnb#1{\lower 2pt\hbox{$\scriptstyle #1$}}
\def\dnc#1{\lower 3pt\hbox{$\scriptstyle #1$}}
\def\upb#1{\raise 2pt\hbox{$\scriptstyle #1$}}
\def\upc#1{\raise 3pt\hbox{$\scriptstyle #1$}}
\def\hprime{\raise 2pt\hbox{$\scriptstyle \prime$}}
\def\ccom{\,\raise2pt\hbox{,}}

\def\asymptotically#1{\;\rlap{\lower 4pt\hbox to 2.0 true cm{
    \hfil\sevenrm  #1 \hfil}}
   \hbox{$\relbar\joinrel\relbar\joinrel\relbar\joinrel
     \relbar\joinrel\relbar\joinrel\longrightarrow\;$}}
\def\Asymptotically#1{\;\rlap{\lower 4pt\hbox to 3.0 true cm{
    \hfil\sevenrm  #1 \hfil}}
   \hbox{$\relbar\joinrel\relbar\joinrel\relbar\joinrel\relbar\joinrel
     \relbar\joinrel\relbar\joinrel\relbar\joinrel\relbar\joinrel
     \relbar\joinrel\relbar\joinrel\longrightarrow$\;}}

\catcode`@=11
\def\C@ncel#1#2{\ooalign{$\hfil#1\mkern2mu/\hfil$\crcr$#1#2$}}
\def\gf#1{\mathrel{\mathpalette\c@ncel#1}}      
\def\Gf#1{\mathrel{\mathpalette\C@ncel#1}}      

\def\gapx{\lower 2pt \hbox{$\buildrel>\over{\scriptstyle{\sim}}$}}
\def\lapx{\lower 2pt \hbox{$\buildrel<\over{\scriptstyle{\sim}}$}}

\def\nablaleft{\hbox{\raise 6pt\rlap{{\kern-1pt$\leftarrow$}}{$\nabla$}}}
\def\nablaright{\hbox{\raise 6pt\rlap{{\kern-1pt$\rightarrow$}}{$\nabla$}}}
\def\nablaboth{\hbox{\raise 6pt\rlap{{\kern-1pt$\leftrightarrow$}}{$\nabla$}}}

\def\boks#1#2{{\hsize=#1 true cm\parindent=0pt   
  {\vbox{\hrule height1pt \hbox
    {\vrule width1pt \kern3pt\raise 3pt\vbox{\kern3pt{#2}}\kern3pt
    \vrule width1pt}\hrule height1pt}}}}

\def\heading{ }
\def\range{ }

\def\body{\vfill\eject\parindent=1.0 true cm
 \ifx\spacing\singlespace\singlespace\else\doublespace\fi}
\def\title#1{\centerline{{\bf #1}}}

\def\today{\ifcase\month\or
  January\or February\or March\or April\or May\or June\or
  July\or August\or September\or October\or November\or December\fi
  \space\number\day, \number\year}
\let\date=\today
\newcount\hour \newcount\minute
\countdef\hour=56
\countdef\minute=57
\hour=\time
  \divide\hour by 60
  \minute=\time
  \count58=\hour
  \multiply\count58 by 60
  \advance\minute by -\count58

\def\sectionskip{\penalty-500\vskip24pt plus12pt minus6pt}

\def\sec{\hbox{\lower 1pt\rlap{{\sixrm S}}{\hbox{\raise 1pt\hbox{\sixrm S}}}}}
\def\section#1\par{\goodbreak\message{#1}
    \sectionskip\nobreak\noindent{\bf #1}\vskip0.3cm \noindent}

\nopagenumbers
\headline={\ifnum\pageno=\count31\frontheadline
  \else{\ifnum\pageno=0\frontheadline
     \else{{\raise 2pt\hbox to \hsize{\paperhead}}}\fi}\fi}

\footline={\centerline{\sevenbf \folio}}

\def\frontheadline{\sevenbf \hfil}
\def\paperhead{\sevenbf \heading\ \range\hfil\folio}
\newdimen\pagewidth \newdimen\pageheight \newdimen\ruleht
\maxdepth=2.2pt
\pagewidth=\hsize \pageheight=\vsize \ruleht=.5pt

\def\onepageout#1{\shipout\vbox{ 
    \offinterlineskip 
  \makeheadline
    \vbox to \pageheight{
         #1 
 \ifnum\pageno=\count31{\vskip 21pt\line{\the\footline}}\fi
 \ifvoid\footins\else 
 \vskip\skip\footins \kern-3pt
 \hrule height\ruleht width\pagewidth \kern-\ruleht \kern3pt
 \unvbox\footins\fi
 \boxmaxdepth=\maxdepth}
 \advancepageno}}

\output{\onepageout{\pagecontents}}

\count31=-1
\topskip 0.7 true cm
\doublespace
\centerline{\bf Gravitational Wave Signatures of Superdense Objects}
\centerline{\bf in a Nonsingular Solution of Gravity Theory}
\centerline{\bf J. W. Moffat}
\centerline{\bf Department of Physics}
\centerline{\bf University of Toronto}
\centerline{\bf Toronto, Ontario M5S 1A7}
\centerline{\bf Canada}
\vskip 0.3 true in
\centerline{\bf Abstract}

A massive gravitationally bound object with a radius $r \leq 2GM/c^2$, which
occurs in the non-singular
version of the nonsymmetric gravitational theory (NGT), replaces the black
hole in Einstein gravity theory.
This object is kept stable by the attractive and repulsive forces
generated by NGT, as well as standard matter pressures, and is called a
superdense object (SDO). The luminosity of gravitational waves emitted
by a SDO with a red-shift of order unity is calculated and it is found that
it could be a strong source of gravitational radiation at low frequencies.
The active galactic nucleus of M87 is
identified with a SDO and the recent observational results obtained by the
Hubble Space Telesope are used to estimate the amplitude of gravitational
radiation.
\vskip 1.5 true in
{\bf UTPT-94-14}
\vskip 0.2 true in
{\bf e-mail: Moffat@medb.physics.utoronto.ca}
\par\vfil\eject
Recently, a class of static spherically symmetric solutions of the nonsymmetric
gravitational theory (NGT) (Moffat 1979) has been studied and found to contain
no event horizons or singularities at $r=0$ (Cornish and Moffat 1994).
A new kind of massive
stable, astrophysical object is found to exist, which replaces
the black hole in Einstein gravity theory (EGT). This object is kept in
hyrodynamic stability by virtue of the attractive gravitational force, and the
repulsive NGT force which supplements the standard matter pressures.
We call this object a ``superdense object'' (SDO).
Newtonian stars, white dwarfs
and neutron stars exist as stable solutions whose hydrodynamic stability
is maintained by means of standard matter equations of state (Cornish 1994).
Since the SDO can be a stable object even for arbitrarily large masses, it can
replace the black hole as a description of very massive and dense
astrophysical systems.

In the following, we shall study SDOs as a strong source of gravitational
waves, and propose that this phenomenon can be used
to observationally distinguish a SDO from a black hole.

The non-singular static spherically symmetric
exterior solution of a SDO is characterized by the four parameters
$M, Q, \ell^2$ and $s$, denoting the mass, electric charge, NGT charge
and a dimensionless parameter, respectively. All the curvature invariants are
finite, and the solution for strong fields depends on $s$ in a non-analytic
way i.e., there is no smooth
limit to EGT for strong fields. This is an important property of
non-singular NGT,
since for even an infinitesimally small value of $s$ there are no
null surfaces in the spacetime and time-like Killing vectors remain time-like
everywhere. Thus, there are no black holes in non-singular NGT. The parameter
$s$ is expected to be a universal constant.

We shall consider for simplicity
SDOs with $Q=0$ and $\ell^2=0$ i.e. electrically and NGT charge neutral
stellar objects. The line element is given in spherical polar coordinates by
$$
ds^2=\gamma dt^2c^2-\alpha dr^2 -r^2(d\theta^2+\hbox{sin}^2\theta d\phi^2).
\numberit
$$
The static spherically symmetric solutions for $\gamma$ and $\alpha$ are
given as expansions in powers
of $m/r$ and $r/m$ for large $r$ and for $r$ near zero, respectively,
with $0 < s < 1$ ($m=GM/c^2$). Solutions for $s > 1$ can be
obtained numerically. For $sm^2/r^2 << 1$, the solution is well
approximated by the Schwarzschild solution.  From this we deduce that
the non-singular solution agrees with the standard experimental tests
of EGT for a suitably chosen value of the constant $s$
(Cornish and Moffat 1994).

The red-shift between $r=0$ and $r=\infty$ is given for $s < 1$ by
$$
z=\hbox{exp}\biggl[{\pi+2s+\pi/8s^2+{\cal O}(s^3)\over 2\vert s\vert}\biggr]-1,
\numberit
$$
while for larger values of $s$, we have
$$
z=\hbox{exp}\biggl[{\pi\over \sqrt{2\vert s\vert}}\biggl(1-
{\pi\vert  s\vert-s(1+2\hbox{exp}(-\pi))+{\cal O}(s^0)\over 2\pi s^2}\biggr)
\biggr] - 1.
\numberit
$$
A calculation of the
stability of neutron stars for $\ell^2=0$ yielded the bound $s \leq 15$
(Cornish 1994). Thus, for physically allowed values, $s\sim 1-7$,
a SDO will be a luminous stellar object with very interesting astrophysical
properties. As $s\rightarrow 0$ the red-shift will increase rapidly and the SDO
will become a dark astrophysical object. We shall assume in the following
that $s$ is in the range $1-7$.

Recently, Hubble space telescope (HST) spectra have been obtained giving
kinematic
evidence for a rotating disk of material in the nucleus of M87
with the radial velocities of the ionized gas measured to be
approximately $\pm 550$ km/sec. This yields a mass for the central object
$\approx 3\times 10^9\,M_{\odot}$ and is taken to be strong but preliminary
evidence for the existence of black holes (Harms {\it et al.} 1994).

The unambiguous theoretical and experimental signature of a black hole would be
the detection of an infinite red-shift event horizon at the
Schwarzschild radius $R_S=2m$. For M87 an estimate of $R_S$ yields:
$$
R_S=8.8\times 10^{14}\,\hbox{cm},
\numberit
$$
which is four orders of magnitude less than the closest point to the center
of M87 that can be observed with the HST. Thus, it is not possible at present
to obtain direct experimental evidence that the nucleus of M87 contains
a black hole.
Moreover, the object contained in the nucleus of M87, given its estimated size
($ > 60$ light yrs), could well be a stable system of stars with a lifetime
$\sim 10^{11}$ yrs.

But there is the alternative possibility that the object is a stable
SDO with a mass $M\sim 3\times 10^9\,M_{\odot}$. Most of the mass will be
concentrated near the center of the SDO. Such an object can possess
very different astrophysical properties from a black hole, which may
allow it to be distiguished experimentally from the latter object.

It is possible that SDOs are formed in the early universe and we call these
objects primordial SDOs. The other possibility is that they are formed by
collapse of a massive star or through accretion of large numbers of stars
in a galaxy. A SDO can be identified with supermassive compact objects
with densities exceeding those of neutron stars. They can also be identified
with large stable, massive objects such as those found in active galactic
nuclei.

The gravitational wave energy radiated by a non-spherical self-gravitating
system is given by (Misner, Thorn and Wheeler 1973; see also Shapiro and
Teukolsky 1983):
$$
E_G\sim Mc^2\biggl({r_G\over R}\biggr)^{7/2},
\numberit
$$
where $M$ and $R$ are the characteristic mass and size of the source,
respectively, and $r_G=GM/c^2$. The power output of gravitational waves
is of the order:
$$
{dE\over dt}\sim {G\over c^5}\biggl({M\over R}\biggr)v^6
\sim \biggl({r_G\over R}\biggr)^2\biggl({v\over c}\biggr)^6 L_0,
\numberit
$$
where
$$
L_0\equiv {c^5\over G}=3.6\times 10^{59}\,\hbox{erg}\,\hbox{sec}^{-1},
\numberit
$$
and $v$ is the velocity of the source. Eliminating $v/c$ from (5), we get
$$
L_{GW}\sim \biggl({r_G\over R}\biggr)^5L_0.
\numberit
$$

We can parametrize (4) and (7) by writing:
$$
\Delta E_G=Mc^2\epsilon^{9/2}\delta^{7/2},\quad \delta={r_G\over R},\quad
\epsilon={\Delta M\over M},
\numberit
$$
and
$$
L_{GW}\sim \delta^5\epsilon^5 L_0,
\numberit
$$
where $\epsilon$ is the mass fraction expected to produce the amplitude
of gravitational radiation.

A gravitational wave will generate a small relative acceleration on test
particles given by
$$
{\ddot \xi}_i={1\over 2}{\ddot h}^{TT}_{ik}\xi_k,
\numberit
$$
where $h^{TT}_{ik}$ denotes the spatial components of the traceless--transverse
weak field symmetric metric.
This produces a small change in the separation of the particle of the size:
$$
\delta\xi_i={1\over 2}h_{ik}^{TT}\xi_k.
\numberit
$$
Then, we have
$$
h\sim {\delta\xi\over \xi} \sim \delta\biggl({r_G\over r}\biggr)\epsilon^2
\sim 1.5\times 10^5\biggl({M/M_{\odot}\over r}\biggr)\epsilon^2.
\numberit
$$

The characteristic time scale for our gravitational wave source is
$$
T\sim \biggl({R^3\over GM}\biggr)^{1/2}\epsilon^{-1/2},
\numberit
$$
and the damping time from gravitational radiation is approximately:
$$
\tau_d\sim {R\over c}\biggl({R\over r_G}\biggr)^3\epsilon^{-3}.
\numberit
$$

Let us assume that the size of the SDO describing the core of M87
is of the order of the Schwarschild radius:
$$
R_S\equiv 2r_G=8.8\times 10^{14}\,\hbox{cm},
\numberit
$$
and that the mass fraction that takes part in the vibration of the SDO
is $\epsilon\sim 10^{-4}$. Then, from (8) we find that
$$
\Delta E_G\sim 5.4\times 10^{45}\,\hbox{erg}.
\numberit
$$
The mean frequency is given by
$$
{1\over T}\sim 2.4\times 10^{-7}\,\hbox{Hz}.
\numberit
$$
and ignoring damping by turbulence, heat conductance and other effects,
the gravitational radiation damping time is
$$
\tau_d\sim 2.3\times 10^{17}\,\hbox{sec}\sim 6\times 10^{10}\,\hbox{periods}.
\numberit
$$

The gravitational wave power output will be of the order:
$$
L_{GW}\sim 1.1\times 10^{38}\,\hbox{erg}/\hbox{sec}.
\numberit
$$
For the approximate distance of M87, $d=1.7\times 10^7$ pc, the flux
measured at Earth will be given by
$$
F\sim 3.2\times 10^{-15}\, \hbox{erg}/\hbox{cm}^2/\hbox{sec}.
\numberit
$$
The fraction of mass of the SDO that produces the gravitational wave amplitude
could be much less than the value we have assumed. The efficiency of the
gravitational wave output would be greater for the collision of two SDOs.

For $\delta \sim 1$, we obtain from (12) for the displacement of the end of
a 10 m bar on the Earth, assuming that $\epsilon\sim 10^{-4}$:
$$
\delta\xi\sim 8.7\times 10^{-17}\,\hbox{cm}.
\numberit
$$
Weber's original bars were sensitive to about $h\sim 10^{-16}$ (Weber 1960) and
new bars will be sensitive to $10^{-20}$ or better, so if the bar or
interferometer is tuned to the low frequency, $\sim 10^{-6}-10^{-7}$ Hz, then
there is a possibility to measure this strong output of gravitational
radiation from the core of an active galactic nucleus. Of course, with
$\delta\sim (0.1-0.01)R_S$ there will be a significant increase in the
gravitational wave flux but with a decrease in the time that the source
produces radiation.

A gravitational wave detector with the sensitivity of the space-based
gravitational wave interferometer LISA/SAGITTARIUS
could see gravitational wave events involving SDOs in the frequency range
$10^{-6}-1$ Hz (Danzmann {\it et al.} 1993, Hellings {\it et al.} 1993).
Earth based detectors would have difficulty
seeing events at such low frequencies due to unavoidable seismic noise.
The question whether a population of supermassive black holes is likely
to emit detactable gravitational waves has been studied in the literature
(see Schutz 1992 for a review). In principle, a significant fraction of the
rest mass energy could be
released during the formation of a black hole, if it involves a
non-axisymmetric
collapse or rotation phase. But calculations have shown that for stellar
mass size sources the emission of gravitational waves would be small
(Rees 1984; Haehnelt 1994).
Another possibility is for a head-on collision of a super massive black hole
with a compact stellar object or the capture of such an object into a
relativistic orbit. But none of these possible sources of black hole
gravitational wave emission are expected to produce as strong a signal
as the SDO, so that the detection of a strong flux of gravitational waves
at a frequency $\sim 10^{-6}-10^{-7}$ Hz could be interpreted as preliminary
evidence for the existence of luminous SDOs.
\par\vfil\eject
{\bf Acknowledgements}
\vskip 0.3 true in
I thank N. J. Cornish, G. Starkmann and S. Tremaine for helpful and stimulating
discussions. This work was supported by the Natural Sciences and Engineering
Research Council of Canada.
\vskip 0.3 true in
\centerline{\bf References}
\vskip 0.3 true in

\noindent Cornish, N. J. 1994, preprint UTPT-94-10.

\noindent Cornish, N. J. and Moffat, J. W. 1994, preprints UTPT-94-4 and
UTPT-94-8.

\noindent Danzmann, K. H. {\it et al.}, 1993, {\it LISA Proposal for a
Laser-Interferometric
Gravitational Wave Detector in Space, report of Max-Planck-Institut f\"ur
Quantenoptik.}

\noindent Haehnelt, M. G. 1994, preprint.

\noindent Harms, R. J. {\it et al.} 1994, preprint.

\noindent Hellings, R. W. {\it et al.} 1993 {\it Jet Propulsion Laboratory
Engineering
Memorandum 314-569}.

\noindent Misner, C. W., Thorne, K. S., and Wheeler, J. A. 1973, {\it
Gravitation}
(San Francisco: W. H. Freeman and Company).

\noindent Moffat, J. W. 1979, {\it Phys. Rev.}, {\bf D19}, 3554.
\item{}. 1979, {\it Phys. Rev.}, {\bf D19}, 3562.
\item{}. 1980, {\it J. Math. Phys.} {\bf 21}, 1978.
\item{}. 1984, {\it Found. Phys.} {\bf 14}, 1217.
\item{}. 1987, {\it Phys. Rev.} {\bf D35}, 3733.
\item{}. 1991, for a review of NGT see, {\it Proceedings of the
Banff Summer Institute on Gravitation, Banff, Alberta, August 12-25, 1990},
eds. by R. Mann and P. Wesson, (Singapore: World Scientific) p.523.

\noindent Moffat, J. W. and Woolgar, E., 1988, {\it Phys. Rev.} {\bf D37}, 918.

\noindent Rees, M. J. 1984, {\it ARA\&A}, {\bf 22}, 471.
\noindent Shapiro, S. L. and Teukolsky, S. A. 1983, {\it Black Holes, White
Dwarfs and Neutron Stars} (New York: John Wiley \& Sons.)

\noindent Shutz, B. F. 1992, {\it First William Fairbank Meeting.} , eds.
Everitt,
C. W.
F., Jantzen, R., Ruffini, R., (Singapore: World Scientific) in press.

\noindent Weber, J. 1960, {\it Phys. Rev.} {\bf D117}, 306.

\end